\documentclass[english,aps,prl,amsmath,twocolumn,preprintnumber,superscriptaddress,notitlepage]{revtex4-1}
\usepackage[T1]{fontenc}
\usepackage[latin9]{inputenc}
\makeatletter

\usepackage{mathptmx,newtxtext,newtxmath,xspace}
\usepackage{amsbsy,bm,bbold}
\usepackage{float}
\usepackage{graphicx,color,xcolor,epsfig,rotate}
\usepackage{fancyhdr}
\usepackage{gensymb}
\usepackage[colorlinks=true,
            linkcolor=blue,
            urlcolor=blue,
            citecolor=blue]{hyperref}
\usepackage{soul}
\setstcolor{red}
\setstcolor{red}

\makeatother
\usepackage{babel}

\begin{document}
\title{Strain-tunable metamagnetic critical endpoint in Mott insulating
rare-earth titanates}
\author{Zhentao Wang}
\affiliation{School of Physics and Astronomy, University of Minnesota, Minneapolis,
Minnesota 55455, USA}
\author{Dominique Gautreau}
\affiliation{School of Physics and Astronomy, University of Minnesota, Minneapolis,
Minnesota 55455, USA}
\affiliation{Chemical Engineering and Materials Science, University of Minnesota,
Minneapolis, Minnesota 55455, USA}
\author{Turan Birol}
\affiliation{Chemical Engineering and Materials Science, University of Minnesota,
Minneapolis, Minnesota 55455, USA}
\author{Rafael M. Fernandes}
\affiliation{School of Physics and Astronomy, University of Minnesota, Minneapolis,
Minnesota 55455, USA}
\date{\today}
\begin{abstract}
Rare-earth titanates are Mott insulators whose magnetic ground state---antiferromagnetic (AFM) or ferromagnetic (FM)---can be tuned
by the radius of the rare-earth element. Here, we combine phenomenology
and first-principles calculations to shed light on the generic magnetic
phase diagram of a chemically substituted titanate on the rare-earth
site that interpolates between an AFM and a FM state. Octahedral rotations
present in these perovskites cause the AFM order to acquire a small
FM component---and vice-versa---removing any multicritical point
from the phase diagram. However, for a wide parameter range, a first-order
metamagnetic transition line terminating at a critical endpoint survives
inside the magnetically ordered phase. Like the liquid-gas
transition, a Widom line emerges from the endpoint, characterized
by enhanced fluctuations. In contrast to metallic FMs, this
metamagnetic transition involves two symmetry-equivalent and insulating
canted spin states. Moreover, instead of a magnetic field, we show
that uniaxial strain can be used to tune this transition to zero temperature,
inducing a quantum critical endpoint.
\end{abstract}

\maketitle

Magnetic quantum phase transitions (QPTs) are often associated with
exotic phenomena~\cite{Sachdev_book}, from strange metallic behavior
to possible deconfined quantum criticality~\cite{Grigera2001,Abanov2003,Senthil2004,Wolfle2007,Ruegg2008,Coldea2010,Jimenez2021}.
Mott insulating perovskites, such as cuprates, iridates, ruthenates,
and titanates, are promising candidates to study such QPTs since
many of the pristine compounds display some type of antiferromagnetic
(AFM) order~\cite{Imada1998,Wen2006}. To suppress the AFM transition
temperature and attempt to induce a QPT, it is often necessary to
dope the compounds, which also favors a metallic state over the Mott
phase.

\begin{figure}
\centering
\includegraphics[width=1\columnwidth]{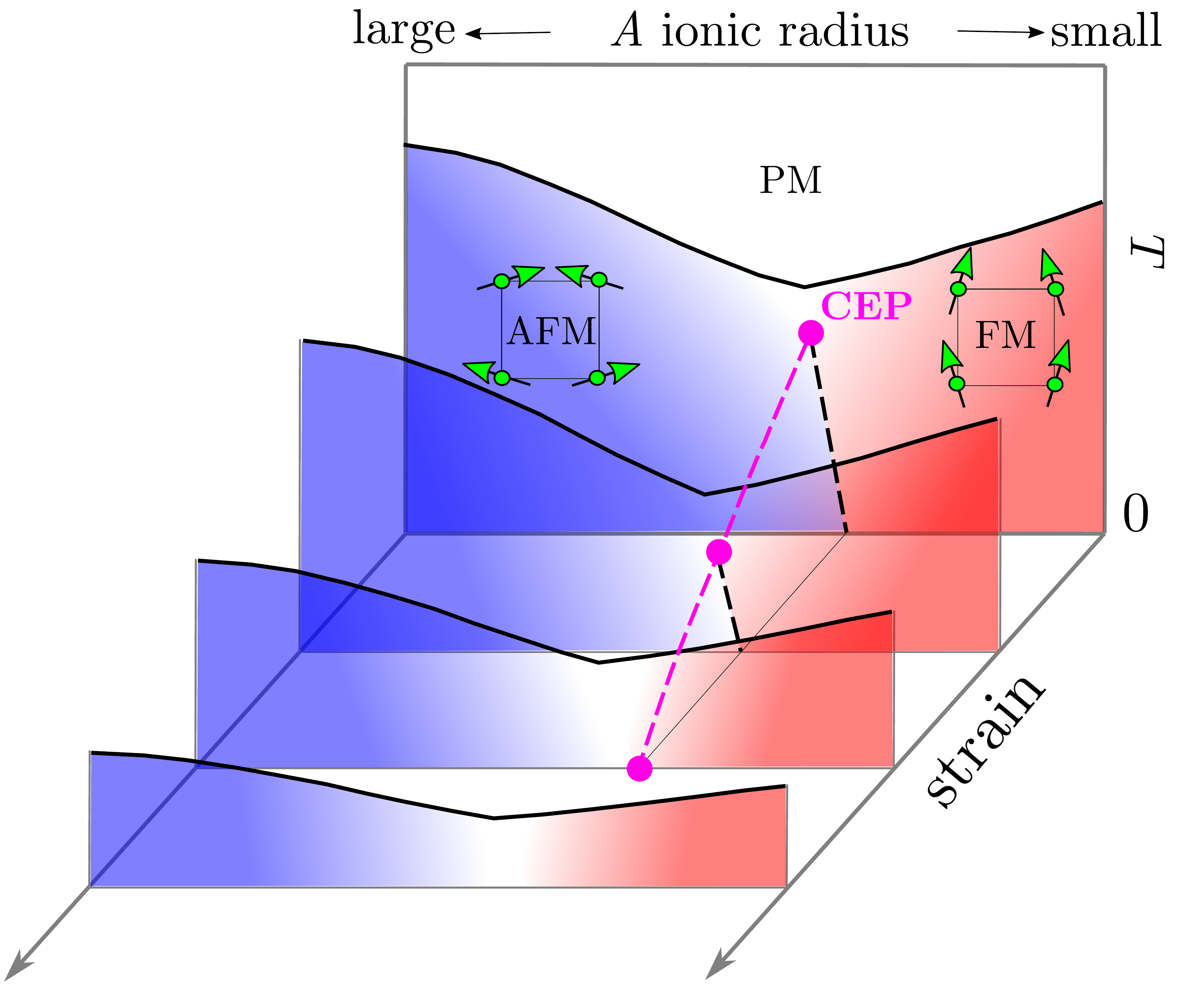}
\caption{Schematic magnetic phase diagram of Mott insulating rare-earth titanates
\emph{A}TiO$_{3}$ as a function of applied uniaxial strain. The horizontal
axis refers to the average rare-earth radius in both stoichiometric
and chemically substituted compounds such as Sm$_{1-x}$Gd$_{x}$TiO$_{3}$
and Y$_{1-x}$La$_{x}$TiO$_{3}$. At low temperatures, the transition
between the antiferromagnetic (AFM)-dominated (blue shaded) and the (ferromagnetic) FM-dominated (red shaded)
states is first order. Due to the TiO$_{6}$ octahedral rotations,
however, the first-order line (black dashed line) terminates at a
critical endpoint (CEP, magenta dot) before reaching the paramagnetic (PM) phase
boundary. The CEP can in principle be tuned to $T=0$ by external
strain, resulting in a quantum critical endpoint (QCEP). The green
arrows show schematically the canted spin configurations in the AFM-
and FM-dominated states. \label{fig:schematic}}
\end{figure}

A remarkable exception are the rare-earth titanates \emph{A}TiO$_{3}$,
where \emph{A} denotes a rare-earth element~\cite{Mochizuki2004b}.
By changing the ionic radius of \emph{A}, the ground state interpolates
between an AFM (of the G type, i.e., a N\'eel AFM) Mott insulating
phase for larger radii (from Sm to La) and a ferromagnetic (FM) Mott
insulating phase for smaller radii (from Yb to Gd)~\cite{Goral1982,Greedan1985,Okimoto1995,Katsufuji1997,Amow2000,ZhouHD2005,Hameed2021},
see Fig.~\ref{fig:schematic}. While the magnetism arises from the
Ti$^{3+}$ $3d^{1}$ state, the size of \emph{A} strongly affects
the rotations of the TiO$_{6}$ octahedra, which in turn impact the
Ti orbital degrees of freedom~\cite{Keimer2009}. The latter are believed
to drive the change from AFM to FM, although the precise mechanism
remains under debate~\cite{Mizokawa1996,Itoh1999,Keimer2000,Khaliullin2000,Buchner2003,Mochizuki2004b,Varignon2017}.

This phenomenology suggests a potential path to realize magnetic QPTs
while remaining inside the Mott insulating state via isovalent chemical substitution
on the rare-earth site~\cite{Choi2015,Goodenough2016}, such as in
Sm$_{1-x}$Gd$_{x}$TiO$_{3}$~\cite{Amow2000} and Y$_{1-x}$La$_{x}$TiO$_{3}$~\cite{Goral1982,Okimoto1995,ZhouHD2005,Hameed2021}.
In principle, there are several possibilities for magnetic QPTs: split
transitions from either the AFM or the FM phase to the paramagnetic
(PM) phase; split transitions from the AFM or FM phases to a coexistent
AFM + FM state; a single first-order AFM-to-FM transition; or a single
second-order AFM-to-FM transition, which would require fine tuning
of parameters.

\begin{figure}
\centering
\includegraphics[width=1\columnwidth]{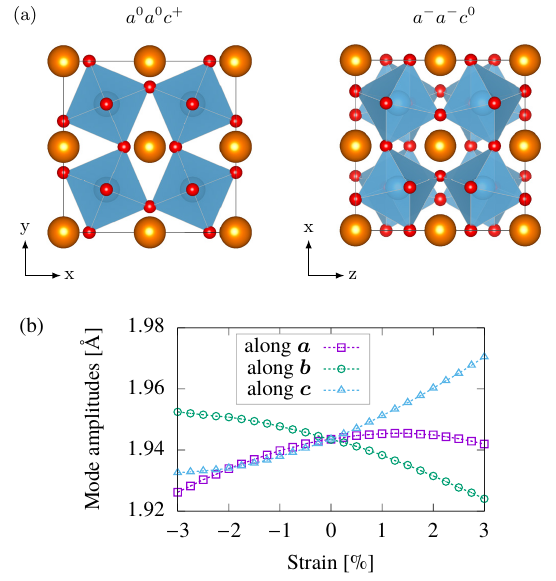}
\caption{(a) Schematics of the two types of octahedral rotation
patterns in rare-earth titanates: in the Glazer notation, they are the $a^{0}a^{0}c^{+}$
pattern (left, which transforms like $M_{2}^{+}$), and the $a^{-}a^{-}c^{0}$
pattern (right, which transforms like $R_{5}^{-}$). Here,
$x$, $y$, and $z$ refer to the cubic axes. (b) Change in the amplitude
of the $R_{5}^{-}$ mode as a function of uniaxial strain along three
different orthorhombic axes of YTiO$_{3}$. }
\label{fig:R_dep}
\end{figure}

There is, however, one important ingredient that qualitatively changes
this scenario: \emph{A}TiO$_{3}$, as several other perovskites, are
not cubic materials but orthorhombic because of the pattern of the
TiO$_{6}$ octahedral rotations {shown in Fig.~\ref{fig:R_dep}(a)}. The corresponding $Pbnm$ crystal
structure generally promotes an admixture between the AFM and FM order
parameters~\cite{Schmitz2005,Bousquet2011}, indicating that most,
if not all, \emph{A}TiO$_{3}$ pristine compounds display the same
mixed AFM + FM phase---a canted spin state, as shown in Fig.~\ref{fig:schematic}.
At first sight, this seems to challenge
the notion that a QPT can be induced via substitution on the rare-earth
site: since the phases of the end compounds are symmetry equivalent,
the only difference is on the relative amplitudes between the AFM
and FM order parameters (and therefore on the canting angle).

In this paper, we combine a phenomenological analysis with first-principles
calculations to show that this is not the case and that a metamagnetic
quantum critical endpoint (QCEP) could still be realized{, particularly in the phase diagrams of Sm$_{1-x}$Gd$_{x}$TiO$_{3}$ and
Y$_{1-x}$La$_{x}$TiO$_{3}$
\cite{Goral1982,Greedan1985,Okimoto1995,Katsufuji1997,Amow2000,ZhouHD2005,Hameed2021}}.
{The key point is that the staggered rotation between neighboring TiO$_{6}$ octahedra, known as the $R_{5}^{-}$ mode and illustrated in the right panel of Fig.~\ref{fig:R_dep}(a), causes an admixture of the AFM and FM order parameters. Microscopically, this effect arises from the changes in the orbital level splittings promoted by the octahedral rotations, which in turn induce magnetic exchange anisotropies as well as the displacement of the oxygen ions from the middle of nearest neighbor titanium pairs, which induces antisymmetric exchange interactions. Because of this admixture, the would-be first-order transition line from the AFM to the FM state splits from the PM phase boundary, as shown schematically in Fig.~\ref{fig:schematic}.}
This first-order line, {denoted by the dashed line in Fig.~\ref{fig:schematic}}, ends in a critical endpoint (CEP, {pink dot in Fig.~\ref{fig:schematic}}), above which
a Widom line emerges, analogous to the liquid-gas phase transition~\cite{Jones1956,Xu2005_Widom,Simeoni2010}.

The relevant order parameter {associated with this transition is not the individual AFM or FM order parameters but the canting angle $\theta$
between the uniform and the staggered magnetizations. As illustrated in Fig.~\ref{fig:schematic}, $\theta$ changes from $\theta\gtrsim0$
deep inside the AFM-dominated phase to $\theta\lesssim\pi/2$ inside the FM-dominated phase}. While $\theta$ changes continuously when the
phase diagram is traversed above the CEP (but below the magnetic ordering
temperature), it undergoes a first-order jump below the CEP{, signaling a metamagnetic transition. Such a metamagnetic transition, driven by octahedral rotations of a Mott insulator and characterized by a jump in the canting angle, is an unexplored counterpart of the widely studied magnetic-field-driven metamagnetic
transition that occurs in metallic FMs, which is signaled by a jump in the uniform magnetization~\cite{Brando2016_RMP}. Remarkably, while the latter can be tuned to $T=0$ by a magnetic field, we show that the CEP uncovered here can be tuned to a QCEP by uniaxial strain. This result is a direct consequence of the sensitivity of the octahedral rotation mode amplitude to strain, which we demonstrate using first-principles calculations [Fig.~\ref{fig:R_dep}(b)]. Because strain has been routinely applied in strongly correlated systems via a variety of experimental setups~\cite{Chu2012,Steppke2017,Sato2017,Pelc2019,Hameed2020}---including recently in rare-earth titanates~\cite{Najev2021}---our results provide a concrete recipe to promote a QCEP in Mott insulating, magnetically ordered perovskites.}


We start by considering the artificial case where there is no spin-orbit
coupling and the octahedral rotations can be neglected, implying a
cubic crystal structure for \emph{A}TiO$_{3}$ (space group $Pm\bar{3}m$).
In this case, the FM and AFM order parameters transform as vectors
under the O(3) spin-rotational group and are denoted respectively
by $\bm{m}$ and $\bm{n}$. The only restriction we impose on the
AFM wave vector $\bm{Q}$ is that $2\bm{Q}=0$, which encompasses
the configurations known as G type: $\bm{Q}=\left(\pi,\pi,\pi\right)$,
C type: $\bm{Q}=\left(\pi,\pi,0\right)$, and A type: $\bm{Q}=\left(0,0,\pi\right)$. In terms of this artificial cubic lattice, they correspond to the
momenta $R$, $M$, and $X$, respectively.
The most general Ginzburg-Landau
(GL) free energy expansion is given by
\begin{equation}
f=\frac{a_{F}}{2}m^{2}+\frac{1}{4}m^{4}+\frac{a_{A}}{2}n^{2}+\frac{1}{4}n^{4}+\frac{\gamma_{1}}{2}m^{2}n^{2}+\frac{\gamma_{2}}{2}\left(\bm{m}\cdot{\bm{n}}\right)^{2}.
\label{eq:F_cubic}
\end{equation}

Here, $a_{F}=a_{F,0}(T-T_{F})$ and $a_{A}=a_{A,0}(T-T_{A})$, where
$T_{F}$ and $T_{A}$ are the bare FM and AFM ordering temperatures,
respectively, and $\gamma_{i}$ are Landau parameters. Note that we
set the positive quartic coefficients of $m^{4}$ and $n^{4}$ to
$1$, which can always be done upon rescaling
the other parameters.
To traverse between the AFM and FM phases, we consider an abstract parameter $\epsilon$, which is a function of the relative concentration of a substituted rare-earth element (as in the cases of Sm$_{1-x}$Gd$_{x}$TiO$_{3}$
and Y$_{1-x}$La$_{x}$TiO$_{3}$) and encompasses not only the structural but also the chemical effects of rare-earth substitution~\cite{Mochizuki2003}. As a result, both $T_{F}$ and $T_{A}$ are implicit functions of $\epsilon$.
The available experimental
phase diagrams of isolavent-substituted rare-earth titanates indicate that $T_{F}$ and $T_{A}$
cross with opposite slopes at $\epsilon^{*}$, i.e., $T_{F}\left(\epsilon^{*}\right)=T_{A}\left(\epsilon^{*}\right)$.
Near this multicritical point, it is convenient to parametrize the
quadratic coefficients as
\begin{equation}
a_{F}=a\left(t-\frac{x}{2}\right),\quad a_{A}=\left(t+\frac{x}{2}\right),\label{eq:parametrization_a}
\end{equation}
where we defined $t\equiv\tilde{T}-\frac{\tilde{T}_{F}+\tilde{T}_{A}}{2}$,
$x\equiv\tilde{T}_{F}-\tilde{T}_{A}$, $a\equiv a_{F,0}/a_{A,0}$,
and $\tilde{T}\equiv a_{A,0}T$. Note that $x$, not to be confused
with doping, is implicitly related to the distance to the multicritical
point as $x\approx\left(\frac{\partial\tilde{T}_{F}}{\partial\epsilon}-\frac{\partial\tilde{T}_{A}}{\partial\epsilon}\right)\left(\epsilon-\epsilon^{*}\right)$.

\begin{figure}
\centering \includegraphics[width=1\columnwidth]{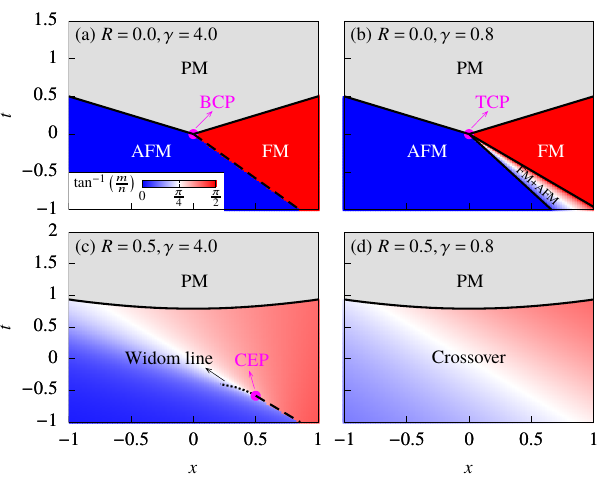}
\caption{Phase diagrams obtained by minimizing the free-energy Eq.~\eqref{eq:GL}
with $a=0.4$. The solid (dashed) lines correspond to second (first)-order transitions. The dotted line in (c) is the Widom line, defined
by the local maxima in the specific heat {[}see Fig.~\ref{fig:cuts}(d){]}.
}
\label{fig:phd}
\end{figure}

A straightforward minimization of Eq.~\eqref{eq:F_cubic} shows that the
nature of the multicritical point depends on the parameter $\gamma\equiv\gamma_{1}+\mathrm{min}\left(\gamma_{2},0\right)$,
being either a bicritical point (BCP) for $\gamma>1$ or a tetracritical
point (TCP) for $\gamma<1$~\cite{Liu1973,Aharony2003}. In the former
case, shown in Fig.~\ref{fig:phd}(a), the AFM and FM phases are
separated by a single first-order transition (dashed line). In
the latter, displayed in Fig.~\ref{fig:phd}(b), there is an intermediate
AFM + FM coexistence phase, separated from the pure AFM and FM phases
by two second-order transitions (solid lines).

We proceed by first turning on the spin-orbit coupling while keeping
the octahedra unrotated. Now, $\bm{m}$ and $\bm{n}$ must transform
according to the magnetic irreducible representations (irreps) associated
with the cubic $Pm\bar{3}m$ space group. The FM and G-type AFM order
parameters can still be parametrized by three-component vectors, as
they transform according to the three-dimensional (3D) irreps $m\Gamma_{4}^{+}$
and $mR_{5}^{-}$ , respectively.
On the other hand, in the case of
C- and A-type orders, the AFM order parameter splits
in a one-dimensional (1D) little-group irrep ($mM_{2}^{+}$ and $mX_{1}^{-}$, respectively) plus a two-dimensional (2D) little-group irrep ($mM_{5}^{+}$ and $mX_{5}^{-}$, respectively) depending on whether
the magnetization is parallel or perpendicular to the wave vectors
$M$ and $X$, respectively.

We are now in position to include the effects of the octahedral rotations
that lower the space group from cubic to the orthorhombic $Pbnm$
(or equivalently $Pnma$ in the standard setting, see Appendix~\ref{sec:group-theory} for details). There are two types of octahedral rotations in the
rare-earth titanates (as in most perovskites), which transform as
the $R_{5}^{-}$ and $M_{2}^{+}$ irreps of the cubic group $Pm\bar{3}m$
[{Fig.~\ref{fig:R_dep}(a)}].
As shown in Appendix~\ref{sec:group-theory}, symmetry makes these modes necessarily couple bilinearly one
component of the FM order parameter $\bm{m} = \left( m_a, m_b, m_c \right)$ to orthogonal
components of two
AFM order parameters  $\bm{n}^{(\mu)} = \left( n^{(\mu)}_a, n^{(\mu)}_b, n^{(\mu)}_c \right)$ (see also Refs.~\cite{Schmitz2005,Bousquet2011}). Here,  \{$a$, $b$, $c$\} denote the $Pbnm$ orthorhombic axes, and $\mu = G, \, A , \, C$ refer to the type of AFM order.
For instance, $m_{c}$ couples bilinearly to both $n_{a}^{(G)}$ and
to $n_{b}^{(A)}$ via combinations of $R_{5}^{-}$ and $M_{2}^{+}$.

These couplings change the form of the
free-energy expansion in Eq.~\eqref{eq:F_cubic}. Hereafter, we focus
on the case where the leading instabilities for $x>0$ and $x<0$
are FM with magnetization along the $c$ axis and G-type AFM with
magnetization along the $a$ axis since this is the common ground
state of most pristine \emph{A}TiO$_{3}$ compounds.
Denoting $m\equiv m_{c}$ and $n\equiv n_{a}^{(G)}$,
the free energy becomes
\begin{equation}
f=\frac{a_{F}}{2}m^{2}+\frac{1}{4}m^{4}+\frac{a_{G}}{2}n^{2}+\frac{1}{4}n^{4}+\frac{\gamma}{2}m^{2}n^{2}-Rmn,\label{eq:GL}
\end{equation}
where $\gamma=\gamma_{1}$ and $R$ is a parameter related to the
strength of the $R_{5}^{-}$-mode octahedral rotations and the spin-lattice coupling~\cite{Birol2012}.


We can now discuss what happens to the phase diagrams of Figs.~\ref{fig:phd}(a) and \ref{fig:phd}(b)
for $R\neq0$ (for concreteness, we consider $R>0$). The first observation is that $m\neq0$ and $n\neq0$
everywhere inside the magnetically ordered phase, resulting in a single
PM-to-magnetic transition and in the disappearance of the multicritical
points. In the case where the unperturbed phase diagram displayed
a TCP and two second-order transition lines ($\gamma<1$), all that
remains is a smooth crossover and no phase transitions {[}Fig.~\ref{fig:phd}(d){]}.
On the other hand, in the case $\gamma>1$, the first-order transition
line detaches from the PM-to-magnetic transition line {[}Fig.~\ref{fig:phd}(c){]}.
While this kills the BCP, it gives rise to a CEP,
located where the first-order line terminates.
The phase diagram of Fig.~\ref{fig:phd}(c)
in the region inside the magnetically ordered state is reminiscent
of the liquid-gas phase diagram of water. While the relevant order parameter in the water case
is the difference in densities of the
two fluid phases, in our magnetic analog it
is the canting of the magnetization, defined as $\theta\equiv\tan^{-1}\left(m/n\right)$
(see Fig.~\ref{fig:phd}).

\begin{figure}
	\centering
	\includegraphics[width=1\columnwidth]{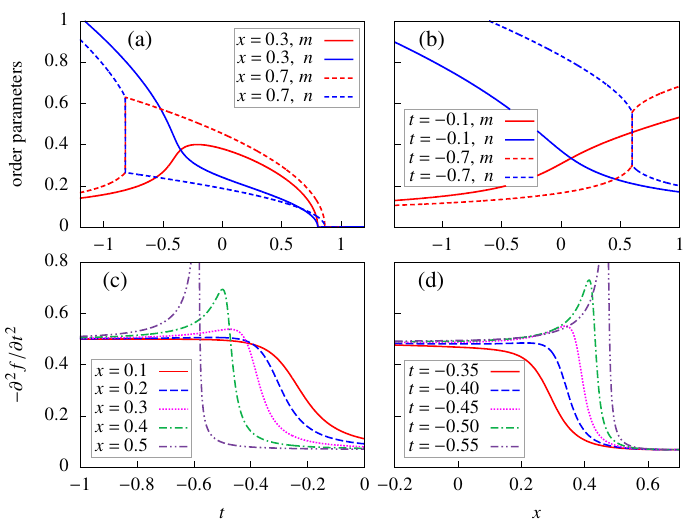}
	\caption{The ferromagnetic (FM) and antiferromagnetic (AFM) order parameters $m$ and $n$ in the phase diagram
		of Fig.~\ref{fig:phd}(c), plotted
		(a) as a function of $t$ for fixed
		$x$ and (b) as a function of $x$ for fixed $t$.
		Specific
		heat $C/T\propto-\partial^{2}f/\partial t^{2}$ (c) as a function of $t$
		for fixed $x$ and (d) as a function of $x$ for fixed $t$.
		The maxima correspond to the Widom line {[}dotted line in Fig.~\ref{fig:phd}(c){]}.}
	\label{fig:cuts}
\end{figure}

To gain a deeper insight, we first plot in Figs.~\ref{fig:cuts}(a) and \ref{fig:cuts}(b)
cuts of the phase diagram of Fig.~\ref{fig:phd}(c) that show how
the order parameters $m$ and $n$ change as a function of $t$ and
$x$ in different regions. Motivated by these results, we introduce a new set of order
parameters $\alpha\equiv(m+n)/2$ and $\beta\equiv m-n$.
Note from Figs.~\ref{fig:cuts}(a) and \ref{fig:cuts}(b) that only $\beta$, but not
$\alpha$, jumps across the first-order line. The free-energy expansion
becomes
\begin{align}
f & =\left(\frac{a_{F}+a_{G}}{2}-R\right)\alpha^{2}+\frac{1+\gamma}{2}\alpha^{4}+\frac{1}{4}\left(\frac{a_{F}+a_{G}}{2}+R\right)\beta^{2}\nonumber \\
 & \quad+\frac{1+\gamma}{32}\beta^{4}+\frac{a_{F}-a_{G}}{2}\alpha\beta+\frac{3-\gamma}{4}\alpha^{2}\beta^{2}.
\end{align}

Therefore, the transition temperature for $\alpha$ ($\beta$) increases
(decreases) due to $R$. Near the CEP, we can eliminate $\alpha$
and obtain a free-energy expansion only in terms of $\beta$,  $f=\sum_{\nu=0}^{4}b_{\nu}\beta^{\nu}$ (see Appendix~\ref{sec:GL} for details).
The first-order line, where $\beta$ jumps, is determined by the condition
that the odd-power coefficients $b_1$ and $b_3$ become zero. Since both are proportional to $a_{F}-a_{G}$, this condition
gives $a_{F}=a_{G}$, leading to
\begin{equation}
x=\frac{2(a-1)}{a+1}t.
\end{equation}

The position of the CEP is given by combining the above expression
with the condition $b_{2}=0$, which yields
\begin{equation}
x_{c}=-\frac{2(a-1)R}{a(\gamma-1)},\quad t_{c}=-\frac{(a+1)R}{a(\gamma-1)}.\label{eq:critical_endpoint}
\end{equation}

Along the first-order line, the order parameter $\beta$ vanishes
according to the usual mean-field result $\left.\beta\right|_{b_{1}=b_{3}=0}=\sqrt{-b_{2}/(2b_{4})}\propto\left(t_{c}-t\right)^{1/2}$.
Beyond a mean-field approximation, however, the CEP is expected to
belong to the Ising universality class. Indeed, by applying the transformation
$x=\frac{2}{a+1}x^{\prime}+(a-1)t^{\prime}$ and $t=\frac{a+1}{2}t^{\prime}$,
the first-order line becomes the vertical line $x^{\prime}=0$ terminating at the CEP at $t_c^{\prime}=-2R/a(\gamma-1)$. Thus, $x^{\prime}$ plays a similar role to the magnetic field in the Ising model, and $t^{\prime}$ to the reduced temperature.
In models described by the Ising universality class, a Widom line
is expected to emerge from the CEP and to extend to higher temperatures~\cite{Jones1956,Xu2005_Widom,Simeoni2010}.
To investigate it, we compute the specific heat $C/T\propto-\partial^{2}f/\partial t^{2}$
as a function of $t$ (or $x$) for fixed $x$ (or $t$). As shown
in Figs.~\ref{fig:cuts}(c) and \ref{fig:cuts}(d), $C/T$ displays a pronounced maximum
above the CEP, indicative of the enhanced fluctuations that characterize
the Widom line~\footnote{The Widom line defined by the maxima in Fig.~\ref{fig:cuts}(c) deviates
slightly from the one defined by the maxima in Fig.~\ref{fig:cuts}(d)
since they are not generally local maxima in the 2D $\{x,t\}$
space.}. As one moves farther from the CEP, this $C/T$ maximum becomes weaker,
and the Widom line fades away, as shown by the dotted line in the
phase diagram of Fig.~\ref{fig:phd}(c).

Our results in Eq.~\eqref{eq:critical_endpoint} show that the temperature of the CEP decreases linearly
with the parameter $R$. Importantly, the $R_{5}^{-}$ rotation mode
is very sensitive to the lattice parameters, which opens the possibility
of tuning the position of the CEP experimentally by applying uniaxial
stress. In Fig.~\ref{fig:R_dep}(b), we show how the $R_{5}^{-}$
mode amplitude changes as a function of uniaxial strain along each
one of the three main orthorhombic axes, determined via first-principles
calculations for the specific case of YTiO$_{3}$ (see Appendix~\ref{sec:dft} for details).
{While} it is difficult
to establish the relative change in $R$ corresponding to the relative
change in the mode amplitude {without input from microscopics}, previous first-principles
calculations show that even modest changes of the order of a few percent
in the amplitudes of the $R_{5}^{-}$ and $M_{2}^{+}$ modes can cause
changes of a factor of $2$ or more in the exchange parameters ~\cite{Najev2021}.
{This} indicates a strong
spin-lattice coupling in rare-earth titanates {and the potential to drastically change the temperature of the CEP via uniaxial strain to promote a QCEP}.
This is also consistent with the strong dependence of the FM transition
temperature on uniaxial pressure observed experimentally in YTiO$_3$~\cite{Keimer2009}.

We now discuss some experimental consequences of our results.
The existence of a first-order line inside the magnetically ordered
state in the chemically substituted rare-earth titanates,
such as Sm$_{1-x}$Gd$_{x}$TiO$_{3}$ and Y$_{1-x}$La$_{x}$TiO$_{3}$,
could be detected by measurements of the temperature dependence of
the canting angle, which would jump across the transition line
and display hysteresis~\footnote{In compounds like Sm$_{1-x}$Gd$_{x}$TiO$_{3}$, the rare-earth atoms possess magnetic moments and order magnetically~\cite{Amow2000,Zhou2005}. Our conclusions remain qualitatively the same if the CEP takes place above the ordering temperature of the rare-earth atoms or when the ordering of the rare-earth atoms is a secondary effect induced by coupling between rare-earth and Ti atoms.}.
 Thermodynamic measurements, such as specific
heat and magnetic susceptibility, should also display typical signatures
of a first-order transition. In the crossover region above the first-order
line, the specific heat is expected to display a pronounced maximum
near the CEP as the Widom line is crossed, indicative of enhanced
fluctuations. While a fine control of the concentration of the substituted
rare-earth may be challenging, one can
alternatively use uniaxial strain as a tuning parameter to smoothly
move a given composition across the first-order and/or Widom lines.

More broadly, for appropriate concentrations, uniaxial strain could
even be used to tune the CEP to zero temperature, thus promoting a
QCEP. The general properties of a QCEP have been discussed in the context
of other materials, most notably quasi-2D organic salts~\cite{Furukawa2015}
and metallic FMs~\cite{Brando2016_RMP}. In the former,
a pressure-induced first-order line separates the PM Mott insulating
and metallic phases, and the transition temperature associated with
the CEP is very low~\cite{Terletska2011,Tremblay2010,Vojta2019}.
Behaviors typically associated with quantum criticality have been
observed in transport properties in the vicinity of the CEP~\cite{Furukawa2015}.
In metallic FMs, an external magnetic field can be used to
suppress the ferro-metamagnetic transition down to $T=0$~\cite{Brando2016_RMP}.
In the particular case of the bilayer ruthenate Sr$_{3}$Ru$_{2}$O$_{7}$,
the putative magnetic-field-driven QCEP is preempted by nematic order
\cite{Borzi2007} and is associated with unusual transport and thermodynamic
properties~\cite{Grigera2001,Millis2002}. Recently, a CEP associated
with a first-order QPT was also observed in a frustrated quantum magnet
\cite{Jimenez2021}. The rare-earth titanates thus provide a
framework to investigate magnetic QCEPs. In contrast to the more usual
case of metallic FMs, the metamagnetic transition involves
two symmetry-equivalent insulating canted spin states, where orthogonal
FM and AFM order parameters coexist. As a result, the dynamical critical exponent of the QCEP is expected to be $z=1$, placing the system at the upper critical dimension since $d+z=4$. Another important difference
is that the metamagnetic transition can be tuned by strain even in
the absence of magnetic fields. Finally, rare-earth titanates can
also be doped with charge carriers, which promotes a Mott insulating-to-metal transition~\cite{Katsufuji1997}. An interesting direction
for future studies is to investigate the fate of the QCEP uncovered
here as the system is tuned across the band-filling-driven Mott transition since a metallic QCEP would have fundamentally different dynamics than an insulating one.

\begin{acknowledgments}
We thank A. Chubukov, M. Greven, S. Hameed, A. Najev, and D. Pelc
for fruitful discussions. This paper was funded by the U.S. Department
of Energy through the University of Minnesota Center for Quantum Materials,
under Grant No. DE-SC-0016371.
\end{acknowledgments}

\appendix




\section{Group-theory analysis}\label{sec:group-theory}
The perovskite rare-earth titanates have space group \#62 due
to the combination of two symmetry-inequivalent rotation patterns
of the TiO$_{6}$ octahedra. As illustrated in Fig.~\ref{fig:R_dep}(a), these rotations transform as the $M_{2}^{+}$
and $R_{5}^{-}$ irreps of the artificial cubic lattice. The pattern
corresponding to the $M_{2}^{+}$ irrep has TiO$_{6}$ octahedra which
rotate about a single cubic axis, while the rotation pattern corresponding
to the $R_{5}^{-}$ irrep is composed of a combination of octahedral
rotations about both of the remaining cubic axes. Using Glazer notation,
we denote these two patterns as $a^{0}a^{0}c^{+}$ and $a^{-}a^{-}c^{0}$,
respectively. Since these rotations lower the cubic symmetry, we are
free to choose any of the three cubic axes as the axis corresponding
to the $M_{2}^{+}$ irrep. In the resulting distorted structure, this
axis points along the longest of the orthorhombic lattice vectors.
If we choose this axis to be $\bm{c}$, as we have here, then the
resulting $a^{-}a^{-}c^{+}$ tilting pattern takes the $Pm\bar{3}m$
cubic symmetry to the $Pbnm$ setting of space group \#62. Had we chosen
the long axis to point along $\bm{b}$ (Glazer tilting pattern $a^{-}c^{+}a^{-}$),
we would have the standard $Pnma$ setting of the same space group.
The basis transformation from the standard setting $Pnma$ to the
non-standard $Pbnm$ is defined by~\cite{Schmitz2005}:
\begin{equation}
\{\bm{a}_{n},\bm{b}_{n},\bm{c}_{n}\}=\{\bm{a}_{s},\bm{b}_{s},\bm{c}_{s}\}P,
\end{equation}
where
\begin{equation}
P=\begin{pmatrix}0 & 1 & 0\\
0 & 0 & 1\\
1 & 0 & 0
\end{pmatrix}
\end{equation}
and the subscripts $s$ and $n$ denote the lattice vectors in the standard and nonstandard settings, respectively.

\begin{figure}
\centering
\includegraphics[width=1\columnwidth]{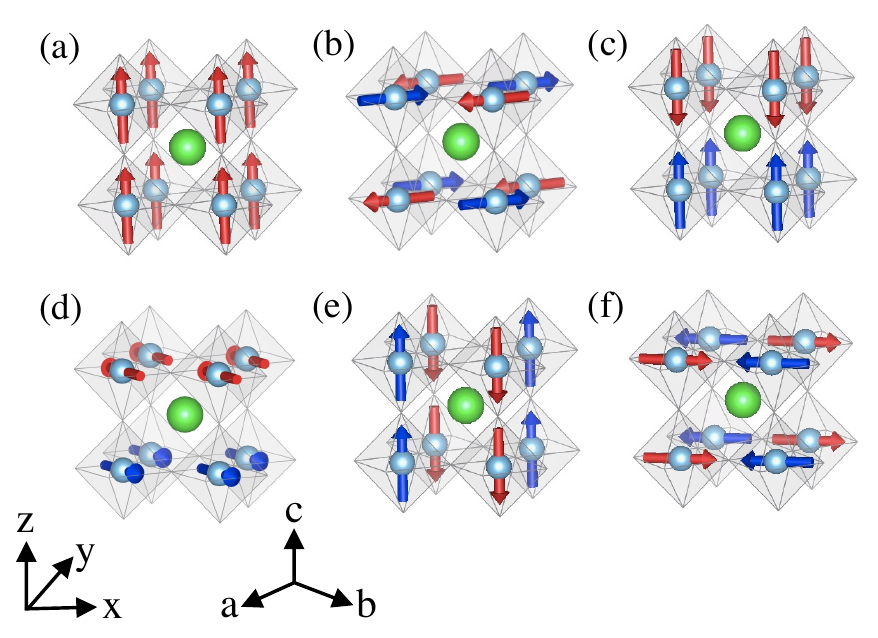}
\caption{The magnetic orders considered in this paper, corresponding to the (a) $m\Gamma_4^+$ (FM order), (b) $mR_5^-$ (G-type AFM order), (c) $mX_1^-$  (A-type AFM order), (d) $mX_5^-$ (A-type AFM order), (e) $mM_2^+$ (C-type AFM order), and (f) $mM_5^+$ (C-type AFM order) irreps of the cubic $Pm\bar{3}m$ space group.
The blue spheres correspond to Ti atoms and the green, to the rare-earth atoms. }
\label{fig:irreps}
\end{figure}

Using standard group-theory techniques and the \textsc{isotropy} Software Suite~\cite{isotropy_site,Stokes2016_isosubgroup},
we determine how the two types of octahedral rotations, $M_{2}^{+}$
and $R_{5}^{-}$, couple the FM order parameter to the three types
of AFM order parameters (G, C, and A type), see also Ref.
\cite{Schmitz2005}. As explained in the main text, the FM order parameter
transforms as $m\Gamma_{4}^{+}$ with respect to symmetry operations of the cubic $Pm\bar{3}m$ space group.
Similarly, the G-type AFM order parameter transforms as $mR_{5}^{-}$. These irreps are 3D since the magnetic moments can point along any of the three axes, and the stars of $\Gamma$ and $R$ contain only one wave vector. The order parameters described by these irreps are shown in Figs.~\ref{fig:irreps}(a) and \ref{fig:irreps}(b).

The A-type AFM order parameters have the wave vector $X=(\pi/a,0,0)$, which has three vectors in its star. For each of these wave vectors, the magnetic moments can point along three orthogonal directions, and as a result, there are nine possible A-type AFM orders. These orders correspond to the 3D irrep $^*mX_{1}^{-}$, in which the magnetic moments align along the wave vector direction, and the six-dimensional (6D) irrep $^*mX_{5}^{-}$, in which the magnetic moments align perpendicular to the wave vector.
We use the $^*$ symbol to differentiate the representations of the space group, which contain all vectors in the star of a wavevector, from the representations of the little group.
For a single wave vector in the star, the magnetic little-group representations $mX_{1}^{-}$ and $mX_{5}^{-}$ are 1D and 2D, respectively.
These modes are depicted in Figs.~\ref{fig:irreps}(c) and \ref{fig:irreps}(d).
Similarly, the C-type AFM orders, which have one of the three wave vectors in the star of $M=(\pi/a, \pi/a,0)$, are described by the magnetic space group irreps $^*mM_{2}^{+}$ (3D) and $^*mM_{5}^{+}$ (6D), which have magnetic little-group irreps $mM_{2}^{+}$ and $mM_{5}^{+}$ that are 1D and 2D, respectively [Figs.~\ref{fig:irreps}(e) and \ref{fig:irreps}(f)].

We consider only the order parameter directions of the above irreps that preserve the translational symmetry of the $Pbnm$ structure. For the FM and G-type AFM orders, all three possible moment directions along the orthorhombic axes preserve the symmetry. Therefore, we denote their respective order parameters as ($F_a, F_b, F_c$) and ($G_a, G_b, G_c$). These correspond to the $m\Gamma_4^+$ and $mR_5^-$ irreps, respectively, and the subscript denotes along which orthorhombic axis the magnetic moments point.

For A- and C-type AFM orders, because only one vector from the star {preserves the translational symmetry of $Pbnm$}, it is sufficient to work with the magnetic little groups.
In the case of A-type AFM order, the $Pbnm$-preserving configurations consist of moments that alternate along the orthorhombic $c$-axis. Therefore, we denote the order parameters $A_c$ and ($A_a, A_b$), associated with the $mX_1^-$ and $mX_5^-$ irreps, respectively. Similarly, the translational symmetry-preserving C-type AFM order parameters have a wave vector on the $a$-$b$ plane and are denoted by $C_c$ and ($C_a$, $C_b$), coming from the $mM_2^+$ and $mM_5^+$ little-group irreps, respectively. We emphasize that, although we have grouped terms that correspond to the same irrep in the high-symmetry cubic $Pm\bar{3}m$ space group, all order parameters listed here are 1D with respect to the lower-symmetry $Pbnm$ space group. That is, no operation in the $Pbnm$ space group transforms $F_a$ to $F_b$, $C_a$ to $C_b$, etc.

\begin{table}[tbp]
\caption{Symmetry-allowed third- and fourth-order terms in the free-energy
expansion that couple the magnetic order parameters with the octahedral
rotation modes in the rare-earth titanates. Note that different subtables
correspond to the different magnetic ground states listed in Table~\ref{table:group62}.}
\begin{centering}
\begin{tabular}{|c|c|}
\hline
Third order & Fourth order\tabularnewline
\hline
$R_{5}^{-}G_{a}F_{c}$ & $R_{5}^{-}M_{2}^{+}A_{b}F_{c}$\tabularnewline
\hline
$M_{2}^{+}G_{a}A_{b}$ & \tabularnewline
\hline
\end{tabular}\quad{}%
\begin{tabular}{|c|c|}
\hline
Third order & Fourth order\tabularnewline
\hline
$R_{5}^{-}F_{a}G_{c}$ & $R_{5}^{-}M_{2}^{+}C_{b}G_{c}$\tabularnewline
\hline
$M_{2}^{+}F_{a}C_{b}$ & \tabularnewline
\hline
\end{tabular}
\par\end{centering}
\begin{centering}
\bigskip{}
\par\end{centering}
\centering{}%
\begin{tabular}{|c|c|}
\hline
Third order & Fourth order\tabularnewline
\hline
$R_{5}^{-}C_{a}A_{c}$ & $R_{5}^{-}M_{2}^{+}F_{b}A_{c}$\tabularnewline
\hline
$M_{2}^{+}C_{a}F_{b}$ & \tabularnewline
\hline
\end{tabular}\quad{}%
\begin{tabular}{|c|c|}
\hline
Third order & Fourth order\tabularnewline
\hline
$R_{5}^{-}A_{a}C_{c}$ & $R_{5}^{-}M_{2}^{+}C_{c}G_{b}$\tabularnewline
\hline
$M_{2}^{+}A_{a}G_{b}$ & \tabularnewline
\hline
\end{tabular}
\label{table:symmetry-allowed}
\end{table}

The symmetry-allowed bilinear couplings between the FM and AFM order
parameters, mediated by the octahedral rotations, are shown in Table
\ref{table:symmetry-allowed}. They result in four different magnetic
ground states that do not break the translational symmetry of the
$Pbnm$ space group and transform as four different $\Gamma$-point
irreps of $Pbnm$~\cite{Schmitz2005}. Table \ref{table:group62}
lists these four ground states in both $Pbnm$ and $Pnma$ notations.
While our analysis in the main text focused on the $G_{a}A_{b}F_{c}$
ground state, with a $G_{a}$-dominated phase crossing over to a $F_{c}$-dominated
phase, the same conclusions would hold for other combinations of FM
and AFM order parameters if they are associated with the
same magnetic ground state.

\begin{table}[btp]
\caption{The four different types of translational-symmetry preserving magnetic
ground states in the different basis settings of the space group \#62.}
\centering %
\begin{tabular}{|c|c|}
\hline
$Pnma$  & $Pbnm$\tabularnewline
\hline
$A_{a}F_{b}G_{c}$  & $G_{a}A_{b}F_{c}$\tabularnewline
\hline
$C_{a}G_{b}F_{c}$  & $F_{a}C_{b}G_{c}$\tabularnewline
\hline
$F_{a}A_{b}C_{c}$  & $C_{a}F_{b}A_{c}$\tabularnewline
\hline
$G_{a}C_{b}A_{c}$  & $A_{a}G_{b}C_{c}$\tabularnewline
\hline
\end{tabular}
\label{table:group62}
\end{table}

\section{First-principles calculations}\label{sec:dft}
To
illustrate how one might experimentally tune the amplitude of the
octahedral rotations, we use first-principles calculations to obtain
the rotation amplitude $R$ as a function of uniaxial strain applied
along the three orthorhombic axes {[}Fig.~\ref{fig:R_dep}(b){]}.
To do this, we fix the strained lattice parameter along an orthorhombic axis to different values while allowing the other two lattice parameters and the internal coordinates of the atoms to relax {to minimize the forces on the atoms and the stresses on the unit cell}. Density functional theory calculations were performed using the projector augmented wave
approach as implemented in the Vienna {\it Ab Initio} Simulation Package
(\textsc{vasp})~\cite{Kresse1993,Kresse1996CMS,Kresse1996PRB}. These calculations
on YTiO$_{3}$ used the PBEsol exchange-correlation functional~\cite{Perdew2008},
a plane-wave cutoff of 550 eV, and a $4\times 4\times 4$ Monkhorst-Pack grid. In
addition, to properly reproduce the local magnetic moments on the
Ti ions, we used the rotationally invariant LSDA + $U$ scheme introduced
by Dudarev {\it et al.} with $U=4$~eV~\cite{Dudarev1998}.

In Fig.~\ref{fig:R_dep}(b), we plot the total displacement amplitude of the $R_5^-$ octahedral rotation mode. While this quantity is proportional to the octahedral rotation angles to first order, it is more well defined. The definition of the octahedral rotation angles can become nonunique when there are multiple different structural distortions that lead to unequal bond lengths, etc., as is the case in YTiO$_3$. The mode amplitude is calculated by projecting the displacements of all atoms to the symmetry-adapted basis modes of the $R_5^-$ irrep and calculating the total magnitude of the projected displacements. We use the \textsc{isodistort} tool~\cite{Campbell2006} for this calculation~\footnote{The details of the procedure can be found in \href{https://stokes.byu.edu/iso/isodistorthelp.php}{https://stokes.byu.edu/iso/isodistorthelp.php}.}.

\section{Ginzburg-Landau expansion near the CEP}\label{sec:GL}
The results of Figs.~\ref{fig:phd} and \ref{fig:cuts} were obtained through
exact numerical minimization. As explained in the main text, to gain
deeper insight into the problem, we find it useful to have an analytical
description of the CEP that appears when $R\neq0$. Expanding the
free-energy in powers of the rotated order parameters $\alpha=(m+n)/2$
and $\beta=m-n$, which correspond to the high-temperature magnetic
transition and the low-temperature first-order transition, respectively,
we find
\begin{align}
f & =\left(\frac{a_{F}+a_{G}}{2}-R\right)\alpha^{2}+\frac{1+\gamma}{2}\alpha^{4}+\frac{1}{4}\left(\frac{a_{F}+a_{G}}{2}+R\right)\beta^{2}\nonumber \\
 & \quad+\frac{1+\gamma}{32}\beta^{4}+\frac{a_{F}-a_{G}}{2}\alpha\beta+\frac{3-\gamma}{4}\alpha^{2}\beta^{2}.
\end{align}

Near the CEP ($\beta\ll1$), we can expand $\alpha$ in powers
of $\beta$:
\begin{equation}
\alpha=\sum_{\nu=0}^{4}\alpha_{\nu}\beta^{\nu}+\mathcal{O}(\beta^{5}),
\end{equation}
where the coefficients $\alpha_{\nu}$ are determined by $\partial_{\alpha}f=0$:
\begin{subequations}
\begin{align}
\alpha_{0} & =\sqrt{-\frac{a_{F}+a_{G}-2R}{2(1+\gamma)}},\\
\alpha_{1} & =\frac{a_{F}-a_{G}}{4\left(a_{F}+a_{G}-2R\right)},\\
\alpha_{2} & =\frac{6(1+\gamma)\alpha_{1}^{2}+\frac{3-\gamma}{2}}{2\left(a_{F}+a_{G}-2R\right)}\alpha_{0},\\
\alpha_{3} & =\frac{2(1+\gamma)\left(\alpha_{1}^{2}+6\alpha_{0}\alpha_{2}\right)+\frac{3-\gamma}{2}}{2\left(a_{F}+a_{G}-2R\right)}\alpha_{1},\\
\alpha_{4} & =\frac{6(1+\gamma)\left(\alpha_{1}^{2}\alpha_{2}+\alpha_{0}\alpha_{2}^{2}+2\alpha_{0}\alpha_{1}\alpha_{3}\right)+\frac{3-\gamma}{2}\alpha_{2}}{2\left(a_{F}+a_{G}-2R\right)}.
\end{align}
\label{eq:alpha's} \end{subequations}

\begin{figure}
\centering
\includegraphics[width=1\columnwidth]{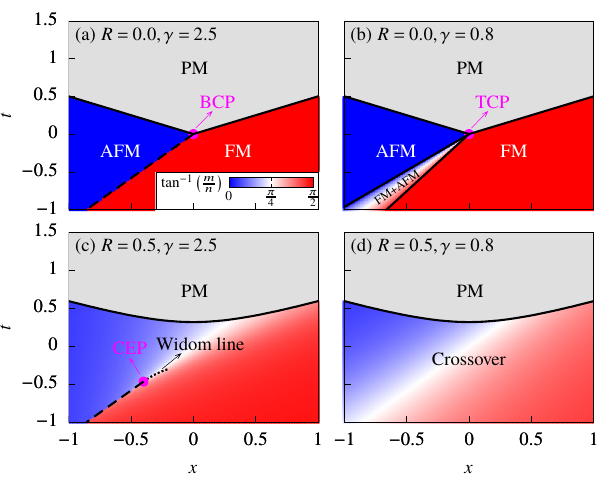}
\caption{Phase diagrams obtained by minimizing the free-energy Eq.~\eqref{eq:GL} with $a=2.5$. The notations follow Fig.~\ref{fig:phd}.
}
\label{fig:S2}
\end{figure}

With Eq.~\eqref{eq:alpha's}, we can eliminate $\alpha$ in
the free-energy, leading to a theory with a single variable $\beta$:
\begin{equation}
f=\sum_{\nu=0}^{4}b_{\nu}\beta^{\nu}+\mathcal{O}(\beta^{5}),
\end{equation}
with \begin{subequations}
\begin{align}
b_{0} & =-\frac{\left(a_{F}+a_{G}-2R\right)^{2}}{8\left(1+\gamma\right)},\\
b_{1} & =\frac{a_{F}-a_{G}}{2}\alpha_{0},\\
b_{2} & =\frac{a_{F}-a_{G}}{4}\alpha_{1}+\frac{\left(\gamma-1\right)\left(a_{F}+a_{G}\right)+4R}{4\left(1+\gamma\right)},\\
b_{3} & =\left[2\left(1+\gamma\right)\alpha_{1}^{2}+\frac{3-\gamma}{2}\right]\alpha_{0}\alpha_{1},\\
b_{4} & =-4\left(1+\gamma\right)\alpha_{1}^{4}-\frac{3-\gamma}{2}\alpha_{1}^{2}+\frac{\gamma-1}{4(1+\gamma)}.
\end{align}
\end{subequations}

The first-order line is obtained by imposing $b_{1}=b_{3}=0$, which
implies $a_{F}=a_{G}$. This leads to the $x(t)$ line defined by
\begin{equation}
x=\frac{2(a-1)}{a+1}t,\label{eq:xt_line}
\end{equation}
where, as in the main text, we used the parametrization
\begin{equation}
a_{F}=a\left(t-\frac{x}{2}\right),\quad a_{G}=t+\frac{x}{2}.
\end{equation}

The CEP is then obtained by combinining $x=\frac{2(a-1)}{a+1}t$ with
$b_{2}=0$, which leads to \begin{subequations}
\begin{align}
x_{c} & =-\frac{2(a-1)R}{a(\gamma-1)},\\
t_{c} & =-\frac{(a+1)R}{a(\gamma-1)}.
\end{align}
\end{subequations}

The order parameter along the first-order line is given by $\partial_{\beta}f=0$:
\begin{equation}
\left.\beta\right|_{b_{1}=b_{3}=0}=\sqrt{-\frac{b_{2}}{2b_{4}}}.
\end{equation}

The parameter $a\equiv a_{F,0}/a_{G,0}$ controls the relative size of the AFM and FM dominated regions but does not affect any other conclusions of this paper. In observation of the experimental phase diagrams where the AFM phase typically occupies a larger area than the FM phase, we used $a=0.4$ in Fig.~\ref{fig:phd}. Figure~\ref{fig:S2} shows the effect of choosing a larger value of $a$, which indeed leads to a larger area occupied by the FM phase.

\bibliographystyle{apsrev4-1}
\bibliography{ref}

\end{document}